%% file: main.tex
\documentclass[conference]{IEEEtran}
\IEEEoverridecommandlockouts
\usepackage{cite}
\usepackage{amsmath,amssymb,amsfonts}
\usepackage{algorithmic}
\usepackage{graphicx}
\usepackage{textcomp}
\usepackage{xcolor}
\usepackage{multirow}
\usepackage{xspace}
\usepackage{hyperref}
\usepackage{xspace}
\usepackage{graphicx}
\usepackage{longtable}
\usepackage{tabularx}
\usepackage{booktabs}
\newcommand*{\CodeGenBase}{\texttt{CodeGen-2B}\xspace}
\newcommand*{\DeepSeek}{\texttt{DeepSeek-Coder-1.3B}\xspace}
\newcommand*{\StarCoder}{\texttt{StarCoder2-3B}\xspace}
\newcommand*{\CodeLlama}{\texttt{CodeLlama-7B}\xspace}

\newcommand*{\CURE}{\texttt{CURE}\xspace}

\usepackage{color, colortbl}
\usepackage[many]{tcolorbox}
\usepackage{comment}

\def\BibTeX{{\rm B\kern-.05em{\sc i\kern-.025em b}\kern-.08em
    T\kern-.1667em\lower.7ex\hbox{E}\kern-.125emX}}

\newcommand{\rqone}{\textbf{RQ$_1$}: \emph{
To what extent does \CURE reduce the generation of deprecated APIs compared to baseline models?
}}

\newcommand{\rqtwo}{\textbf{RQ$_2$}: \emph{
How effectively does \CURE promote correct API replacements compared to suppression-based unlearning approaches?
}}

\newcommand{\rqthree}{\textbf{RQ$_3$}: \emph{
How does \CURE impact general code generation performance on standard benchmarks?
}}

\newtcolorbox{shadedbox}{
	drop shadow southeast,
	breakable,
	enhanced jigsaw,
	colback=white,
	boxrule=0.80pt,
	left=0.3em,
	right=0.3em,
	top=0.1em,
	bottom=0.05em
}

\begin{document}

\title{From Forgetting to Replacing:
Contrastive Unlearning for Handling Deprecated APIs in Code LLMs}

\title{Towards Knowledge Alignment in Code LLMs:
Contrastive Unlearning for Evolving APIs}

%\author{Anonymous authors}

\author{
\IEEEauthorblockN{Huy Q. Tran}
\IEEEauthorblockA{%\textit{DISIM} \\
\textit{Hanoi University of Science and Technology (HUST)}\\
Hanoi, Vietnam\\
Huy.TQ226109@sis.hust.edu.vn
}
\and
\IEEEauthorblockN{Dang H. Vu}
\IEEEauthorblockA{%\textit{DISIM} \\
\textit{HUST}\\
Hanoi, Vietnam\\
dang.vh225962@sis.hust.edu.vn
}
\and
\IEEEauthorblockN{Tuyen N.	Dinh}
\IEEEauthorblockA{%\textit{DISIM} \\
\textit{HUST}\\
Hanoi, Vietnam\\
tuyen.dn235242@sis.hust.edu.vn
}
\and
\IEEEauthorblockN{Anh H. D.	Nguyen}
\IEEEauthorblockA{%\textit{DISIM} \\
\textit{HUST}\\
Hanoi, Vietnam\\
Anh.NDH2416662@sis.hust.edu.vn	
}
\and
\IEEEauthorblockN{Anh N. H.	Vu}
\IEEEauthorblockA{%\textit{DISIM} \\
\textit{HUST}\\
Hanoi, Vietnam\\
Anh.VHN225471@sis.hust.edu.vn
}
\and
\IEEEauthorblockN{Anh M. T. Bui$^{*}$\thanks{*Corresponding author}}
\IEEEauthorblockA{%\textit{DISIM} \\
\textit{HUST}\\
Hanoi, Vietnam \\
anhbtm@soict.hust.edu.vn}
%\and
%\IEEEauthorblockN{Huy N. D. Pham}
%\IEEEauthorblockA{%\textit{DISIM} \\
%\textit{AI4Life, Hanoi University of Science and Technology}\\
%Hanoi, Vietnam \\
%huypnd0305@gmail.com}
\and
\IEEEauthorblockN{Phuong T. Nguyen}
\IEEEauthorblockA{%\textit{DISIM} \\
\textit{University of L'Aquila}\\
L'Aquila, Italy \\
phuong.nguyen@univaq.it}
}

\maketitle

\begin{abstract}
% Large Language Models (LLMs) have recently shown significant advances in software engineering tasks, particularly in code generation.
% However, they often demonstrate suboptimal utilization of correct and up-to-date Application Programming Interfaces (APIs) due to knowledge cut-off and the rapid evolution of software libraries, frequently generating deprecated API usages that lead to unreliable and incompatible code. 
% Fine-tuning-based methods lack selectivity and are often impractical when only a small portion of the model’s knowledge requires modification. 
Large Language Models (LLMs) have recently achieved strong performance in code generation. However, due to knowledge cut-off and the rapid evolution of software libraries, they often generate deprecated API usages that lead to unreliable and incompatible code. Existing fine-tuning methods lack selectivity when only a small portion of model knowledge requires modification.
Recent model-level approaches, such as machine unlearning and model editing, offer a promising direction for modifying parametric knowledge.
However, their use for deprecated API mitigation remains largely unexplored. Moreover, existing methods primarily suppress outdated APIs, but do not explicitly steer models toward correct replacements, often leading to mismatched or incomplete generations.
To address this limitation, we developed %propose 
\CURE, a contrastive unlearning approach that shifts unlearning from purely suppressing outdated knowledge to explicitly promoting correct API replacements. Concretely, \CURE jointly discourages deprecated APIs while encouraging their valid alternatives, enabling more reliable adaptation to evolving software libraries. The %conducted 
experiments on recent deprecated API benchmark dataset show that \CURE not only reduces deprecated API usage but also improves correct API replacement, while preserving general code generation performance. \CURE substantially outperforms two SOTA baselines with respect to different quality metrics. These findings highlight the importance of combining suppression with replacement when adapting LLMs to evolving software ecosystems.
\end{abstract}


\section{Introduction}
\label{sec:intro}
\input{src/introduction}

\section{Related work}
\label{sec:rw}

\input{src/related-work}

\section{Proposed Method}
\label{sec:proposed}
\input{src/proposed}

\section{Experiment Settings}
\label{sec:settings}
\input{src/exp-settings}

\section{Results and Discussion}
\label{sec:result}
\input{src/results}

\section{Threats to Validity}
\label{sec:threats}
\input{src/threats}

\section{Conclusion}
\label{sec:con}
In this paper, we presented \CURE, a contrastive unlearning approach for mitigating deprecated API usage in Code LLMs. Unlike conventional unlearning methods that primarily suppress undesirable knowledge, \CURE explicitly promotes valid API replacements through a contrastive objective that distinguishes deprecated APIs from their correct alternatives. Initial experimental results show that \CURE effectively reduces deprecated API generation while improving replacement adoption across multiple Code LLMs. Moreover, evaluation on \texttt{HumanEval} indicates that the proposed approach preserves general code generation capability with only minor degradation. These early findings suggest that combining suppression and replacement-aware learning is a promising direction for adapting Code LLMs to evolving software ecosystems.

\section*{Acknowledgment} 
	%A part of the numerical simulations/evaluations has been realized on the Linux HPC cluster Caliban of the High-Performance Computing Laboratory of the Department of Information Engineering, Computer Science and Mathematics (DISIM) at the University of L'Aquila. 
	This paper has been partially supported by the MOSAICO project (Management, Orchestration and Supervision of AI-agent COmmunities for reliable AI in software engineering) that has received funding from the European Union under the Horizon Research and Innovation Action (Grant Agreement No. 101189664). %The work has been partially supported by the EMELIOT national research project, which has been funded by the MUR under the PRIN 2020 program (Contract 2020W3A5FY). 
	%The work has been also partially supported by the European Union--NextGenerationEU through the Italian Ministry of University and Research, Projects PRIN 2022 PNRR \emph{``FRINGE: context-aware FaiRness engineerING in complex software systEms''} grant n. P2022553SL. We acknowledge the Italian ``PRIN 2022'' project TRex-SE: \emph{``Trustworthy Recommenders for Software Engineers,''} grant n. 2022LKJWHC. %We thank the anonymous reviewers for their

\bibliographystyle{IEEEtran}
\bibliography{refs}

\end{document}

%% file: src/introduction.tex
% Large Language Models (LLMs) have recently achieved significant progress in software engineering tasks, particularly in code generation and code completion~\cite{mu2024clarifygpt,bouzenia2025repairagent,harman2025mutation,wang2025can,he2026llm}. 
% Their capabilities are largely enabled by pretraining on massive code corpora collected from public repositories~\cite{fan2023large}. 
% However, this training process also embeds outdated patterns in the model’s parametric memory, including deprecated Application Programming Interfaces (APIs) that remain prevalent in historical code~\cite{wang2025llms}. 
%As a result, Code LLMs may continue to generate obsolete API calls that no longer align with current software ecosystems, leading to reduced reliability and maintainability of the generated code.
Large Language Models (LLMs) have recently shown strong performance in software engineering tasks, particularly code generation and code completion~\cite{mu2024clarifygpt,bouzenia2025repairagent,harman2025mutation,wang2025can,he2026llm}.
These capabilities are enabled by pretraining on massive public code corpora~\cite{fan2023large}, which also exposes models to outdated coding patterns, including deprecated APIs frequently appearing in historical repositories~\cite{wang2025llms}.
As a result, Code LLMs may continue to generate obsolete API calls that no longer align with modern codebases, reducing code reliability.
%This issue arises not only from knowledge cutoff, but also from the way outdated API usage becomes embedded in the model's parameters. 

This issue stems not only from knowledge cutoff, but also from outdated patterns embedded in the model’s parametric memory~\cite{zhu2025your}, which can bias Code LLMs toward deprecated APIs even when valid alternatives exist. In rapidly evolving libraries, such behavior may introduce compatibility issues and increase maintenance effort~\cite{zheng2025humanevo,kuhar2025libevolutioneval}. Although fine-tuning can update model knowledge, it is often inefficient for localized API-level modifications since it globally updates model parameters. The key challenge is therefore to selectively remove outdated API knowledge while preserving the model’s general code generation capability~\cite{li2024wmdp}.

Existing efforts to mitigate outdated API usage follow two main directions. 
Inference-time strategies, e.g., %such as 
prompt augmentation and output-level API replacement~\cite{wang2025llms}, attempt to steer model outputs toward updated APIs without modifying the %model’s 
internal knowledge. However, these approaches assume that each input context can be accurately mapped to a specific replacement API, which may not be realistic in %diverse 
real-world %coding 
scenarios. 
%However, these approaches operate at the surface level and leave outdated knowledge in the model’s parametric memory largely unchanged.
More direct model-level approaches, including model editing~\cite{li2025model,lin2025lightweight} and machine unlearning~\cite{fan2410simplicity,jiang2026large,chu2025scrub}, modify the model's parametric memory to suppress undesirable knowledge.
However, their application to deprecated API mitigation remains limited.
%Lin et al.~\cite{lin2025lightweight} propose an instance-level approach based on dynamic model editing, where the model is updated whenever a deprecated API is generated and then re-evaluated to eliminate the undesired behavior. 
%While effective for correcting individual instances, this approach operates locally and may not consistently generalize across unseen inputs.
Lin et al.~\cite{lin2025lightweight} propose a dynamic %model 
editing approach that updates the model whenever a deprecated API is generated. %However, 
This %instance-level 
strategy operates locally and may not generalize well to unseen contexts.
Recent %unlearning 
approaches~\cite{jiang2026large,chu2025scrub} % for code generation, 
%e.g., %such as
%\texttt{PROD}~\cite{jiang2026large} and \texttt{CodeEraser}~\cite{chu2025scrub}, 
mainly focus on suppressing undesirable knowledge, without explicitly guiding the model toward correct API replacements required for evolving %software 
libraries.

In this paper, we propose \CURE, a \underline{\bf C}ontrastive \underline{\bf U}nlearning approach for deprecated API \underline{\bf RE}placement. 
The key idea is to shift unlearning from a purely suppressive process to a replacement-oriented mechanism. 
Specifically, \CURE builds upon a parametric machine unlearning method, such as \texttt{PROD}~\cite{jiang2026large} or \texttt{SimNPO}~\cite{fan2024simplicity}, to reduce the likelihood of generating deprecated APIs. 
On top of this, we introduce a contrastive objective that explicitly promotes correct API alternatives, enabling the model not only to forget outdated APIs but also to adopt valid replacements.
% We evaluate the effectiveness of our approach on a recent benchmark for deprecated API usage, across four representative Code LLMs including \CodeGenBase, \DeepSeek, \StarCoder and \CodeLlama.
% The experimental results show that \CURE not only reduces the generation of outdated APIs but also significantly improves the accuracy of selecting correct replacements compared to standard unlearning methods.
% To further assess generalization, we also evaluate the model on the \texttt{HumanEval} benchmark~\cite{zheng2025humanevo}. 
% The results indicate that our approach preserves the model’s general code generation capability, without introducing noticeable degradation on standard programming tasks.
We evaluate \CURE on a recent deprecated API benchmark across four representative Code LLMs: \DeepSeek~\cite{guo2024deepseek}, \StarCoder~\cite{lozhkov2024starcoder}, \CodeLlama~\cite{roziere2023code} and \CodeGenBase~\cite{nijkamp2023codegen2}. Experimental results show that \CURE not only reduces deprecated API generation, but also improves correct API replacement compared to standard unlearning methods. Evaluation on \texttt{HumanEval}~\cite{chen2021evaluating} further indicates that \CURE %the proposed approach 
preserves general code generation capability without noticeable degradation.

The main contributions of this work are as follows:

\begin{itemize}
    \item We propose \CURE, a contrastive unlearning framework for mitigating deprecated API usage in Code LLMs through replacement-oriented learning.
    
    \item We introduce a contrastive objective that explicitly aligns deprecated APIs with their corresponding updated alternatives under the same code context.
    
    \item Experiments on a recent deprecated API benchmark and \texttt{HumanEval}~\cite{chen2021evaluating} show that \CURE consistently reduces deprecated API generation, improves correct API replacement, and preserves general code generation capability compared to standard unlearning methods.
    
    \item We release our replication package to support future research~\cite{icsme2026}.
\end{itemize}

%% file: src/related-work.tex
Modern software systems evolve continuously, causing APIs to be frequently revised and deprecated in favor of improved alternatives~\cite{10.1016/j.infsof.2017.09.007}. This is particularly challenging for Code LLMs, whose static training corpora often encode outdated API knowledge, leading to deprecated API generation. 
Wang et al.~\cite{10.1109/ICSE55347.2025.00245} report that 37.4\% of API predictions produced by GPT-3.5 involve deprecated APIs. Existing solutions mainly follow two directions: inference-level methods such as \texttt{REPLACEAPI} and \texttt{INSERTPROMPT}~\cite{10.1109/ICSE55347.2025.00245}, modifying outputs without updating model, % knowledge, 
and model-level approaches including model editing~\cite{lin2025lightweight} and reinforcement learning~\cite{wu2026recode}, which directly adapt knowledge within the model.

Closely related to these efforts, machine unlearning has recently emerged as a promising direction for removing undesirable knowledge from LLMs~\cite{cao2015towards}. Existing approaches can be broadly categorized into exact and approximate unlearning~\cite{chu2025scrub,jang2023knowledge}. While exact methods retrain models from scratch to ensure complete data removal, approximate methods provide a more practical alternative through gradient-based parameter updates, including GA~\cite{jang2023knowledge}, DPO-based unlearning~\cite{rafailov2023direct}, NPO~\cite{zhang2024negative}, and SimNPO~\cite{fan2024simplicity}. More recently, \texttt{PROD}~\cite{jiang2026large} introduces token-level suppression for precise code unlearning, while \texttt{CODEEraser}~\cite{chu2025scrub} selectively removes sensitive memorized code segments. Despite these advances, applying unlearning techniques to mitigate deprecated API usage in Code LLMs remains largely underexplored.

%% file: src/proposed.tex
%Figure~\ref{fig:framework} illustrates the overall framework of our method, which consists of two sequential phases. 
In this section, we first formulate the unlearning problem for mitigating deprecated APIs, then present our proposed approach \CURE, built upon a gradient-based parametric unlearning method by incorporating a contrastive objective 
%This objective is designed to not only suppress outdated API knowledge, but also explicitly 
to promote correct replacements by contrasting deprecated and valid API usages under the same context. 

\subsection{Problem Formulation}
\label{sec:formulation}
We formulate unlearning as a post-training procedure that aims to reduce the model’s tendency to generate deprecated APIs by modifying its parametric memory. Let $\pi_{\theta}$ denote a language model parameterized by $\theta$.
We denote $\mathcal{D}_{f} = \{x_i, y^{-}_i\}_{i=1}^{N}$ as the forgetting set, which consists of code samples associated with deprecated APIs. 
Each pair $(x_i, y_i^-)$ includes a prompt $x_i$ and its corresponding undesirable completion $y_i^-$, where $y_i^-$ contains deprecated API usage. 
In this work, $x_i$ serves as a probing %code 
snippet under the code completion setting, where the model is required to generate the continuation.
The objective of unlearning is to update the model parameters from $\theta$ to $\theta^*$ such that the updated model $\pi_{\theta^*}$ reduces the likelihood of generating undesirable outputs $y_i^-$ given $x_i$. This can be formulated as the following optimization problem:
\begin{equation}
    \theta^* = \text{arg}\min_{\theta} \sum\limits_{i=1}\limits^{N} \mathcal{L}_{unl}(\pi_{\theta}(x_i), y_i^{-})
    \label{eq:unlearning-loss}
\end{equation}
where $\mathcal{L}_{unl}$ denotes an unlearning loss that penalizes the model for generating undesirable outputs.
% As in this work, we not only target to forget deprecated API usages but also to encourage valid alternatives under the same context. 
% We define for each $y_{i}^{-}$, an alternative with the correct API replacement $y_{i}^{+}$ with respect to $x_{i}$.
% To promote correct replacement we incorporate a preference loss $\mathcal{L}_{pref}(\Phi_{\theta}(x_i), y_i^+)$. Equation~\ref{eq:unlearning-loss} is updated as follows.
% \begin{equation}
%     \theta^* = \min_{\theta} \sum\limits_{i=1}\limits^{N} -\mathcal{L}_{unl}(\Phi_{\theta}(x_i), y_i^{-}) + \lambda \mathcal{L}_{pref}(\Phi_{\theta}(x_i), y_i^{+})
%     \label{eq:contrastive-loss}
% \end{equation}
% where $\lambda$ is a regulation term.
%In this work, $x_{f}^{(i)}$ represents a probing code snippet, and we focus on the code completion task in which the model is required to complete the given snippet.

\subsection{Unlearning Instantiations}
\label{sec:unlearning-methods}

The formulation in Equation~\ref{eq:unlearning-loss} defines a general unlearning objective. In practice, different methods instantiate the unlearning loss $\mathcal{L}_{unl}$ in different ways. 
In this work, we consider two representative gradient-based parametric unlearning approaches, \texttt{SimNPO}~\cite{fan2024simplicity} and \texttt{PROD}~\cite{jiang2026large}, which have been shown effective in modifying model behavior at the sequence and token levels, respectively.

\smallskip
\noindent
$\triangleright$ {\bf Simple Negative Preference Optimization.}
\texttt{SimNPO}~\cite{fan2024simplicity} is a sequence-level unlearning method that suppresses undesirable outputs through preference-based optimization. Given a forgetting sample $(x, y^-)$, where $y^-$ includes deprecated API usage, \texttt{SimNPO} suppresses the conditional likelihood of the undesirable sequence while applying length normalization to avoid biasing the optimization toward longer responses.
\begin{equation}
\mathcal{L}_{unl}^{\texttt{SimNPO}}
= -\frac{2}{\beta}
\log \sigma \left(
-\frac{\beta}{|y^-|}
\log \pi_{\theta}(y^-|x)
\right)
\label{eq:simnpo-loss}
\end{equation}
where $\pi_{\theta}$ denotes the current model, $|y^-|$ represents the length of the undesirable response, $\beta$ is a scaling factor controlling the optimization strength, and $\sigma(\cdot)$ denotes the \texttt{sigmoid} function. Intuitively, Eq.~\ref{eq:simnpo-loss} incorporates length normalization via $\frac{1}{|y^-|}$ to mitigate the tendency of longer responses to dominate the optimization through accumulated log-probabilities. 
%This enables stable unlearning while suppressing the generation of deprecated API usages and preserving general generation behavior.

\smallskip
\noindent
$\triangleright$ {\bf Probabilistic Redistribution for Output Distribution.} 
\texttt{PROD}~\cite{jiang2026large} is a token-level %unlearning 
method that suppresses undesirable %code 
snippet generations by directly modifying the target token distribution during training. 
Given a forgetting sample $(x, y^-)$, \texttt{PROD} minimizes the divergence between the current model distribution $\pi_{\theta}$ and a modified target distribution $p_{T}$.
%To achieve surgical precision in source code unlearning-a critical requirement for eliminating memorized deprecated APIs without compromising the syntactic validity of the generated code—we utilize \texttt{PROD}. 
%This method intervenes directly at the token-level output distribution, optimizing the model to match a meticulously sculpted target distribution $p_T$ through cross-entropy loss:
\begin{equation}
\mathcal{L}^{\texttt{PROD}}_{unl} = -\sum_{t=1}^{L} \sum_{w \in \mathcal{V}} p_T(w|x, y^-_{<t}) \log \pi_\theta(w|x, y^-_{<t})
\label{eq:prod-loss}
\end{equation}
where $L$ denotes the sequence length, $\mathcal{V}$ is the vocabulary space, $p_T(\cdot)$ denotes the redistributed target distribution at decoding step $t$, and $\pi_\theta(\cdot)$ represents the probability assigned by the current model to token $w$ given the context $x$ and previously generated tokens $y_{<t}$. Intuitively, the objective function suppresses the probabilities of deprecated API tokens while redistributing probability mass over the remaining vocabulary, thereby discouraging obsolete API usage in generated code.
% In this formulation, the loss $\mathcal{L}_{\texttt{PROD}}$ minimizes the divergence between the current model distribution $\pi_\Theta$ and the target distribution $p_T$. Here, $w \in \mathcal{V}$ denotes a candidate token within the vocabulary, $x_f$ represents the input context, and $y_f$ signifies the specific deprecated code snippet targeted for erasure. The \texttt{PROD} pipeline executes three distinct steps: (1) suppressing the probabilities of tokens in undesirable snippets to zero; (2) eliminating long-tail noise through nucleus sampling; and (3) redistributing the remaining probability mass across the safe vocabulary to faithfully preserve the original statistical patterns of the programming language.

% To enhance the forgetting efficacy for shorter sequences, an $\alpha$-suppression mechanism is applied to the supervisory signal:
% \vspace{-1em}
% \begin{equation}
% p_T(w | x_f, y_{f,<t}) = \begin{cases} -\alpha p_o(w), & \text{if } w = y_{f,t} \\ \tilde{p}_t(w), & \text{otherwise} \end{cases}
% \end{equation}

% where $p_o$ is the original distribution, $\tilde{p}_t$ is the noise-trimmed distribution, and $\alpha$ is a hyperparameter controlling the degree of forgetting. This precise token-level manipulation ensures the thorough erasure of deprecated APIs while preserving the structural scaffolding required for accurate, general-purpose code generation.

\subsection{Contrastive Unlearning for API Replacements}
\label{sec:contrastive}
While the methods in Section~\ref{sec:unlearning-methods} effectively suppress deprecated API usage, they do not explicitly guide the model toward correct replacements. To overcome this limitation, we propose a contrastive unlearning formulation that jointly suppress outdated APIs and promotes valid alternatives under the same context.
For each forgetting sample $(x, y^-)$, we construct a corresponding positive sample $y^+$ by applying API-level transformations based on updated documentation or curated mappings between deprecated APIs and their recommended replacements. This ensures that $(x, y^+)$ represents a desirable completion that preserves the original intent while using up-to-date APIs.
Given a triplet $(x, y^-,y^+)$, we extend the unlearning objective by introducing a contrastive loss that enforces a relative preference for $y^+$ over $y^-$ as follows. %The loss is defined as in Equation~\ref{eq:contrastive-loss}.
% \begin{equation}
%     \mathcal{L}_{con} = -\log \sigma \big(
% \frac{1}{|y^+|}\sum_t \log P(y^+_t \mid x, y^+_{<t}) - \frac{1}{|y^-|}\sum_t \log P(y^-_t \mid x, y^-_{<t}) \big)
%     \label{eq:contrastive-loss}
% \end{equation}
\begin{equation}
\begin{aligned}
\mathcal{L}_{ctr} 
&= -\log \sigma \Big(
    \frac{1}{|y^+|}\sum_t \log P(y^+_t \mid x, y^+_{<t}) \\
&\qquad\quad\quad
    - \frac{1}{|y^-|}\sum_t \log P(y^-_t \mid x, y^-_{<t})
\Big)
\end{aligned}
\label{eq:contrastive-loss}
\end{equation}
The overall objective is defined as in Equation~\ref{eq:overall-loss}.
\begin{equation}
    \mathcal{L}_{\texttt{CURE}} = \mathcal{L}_{unl} + \lambda \mathcal{L}_{ctr} 
    \label{eq:overall-loss}
\end{equation}
where $\lambda$ controls the trade-off between suppressing deprecated APIs and promoting correct replacements, $\mathcal{L}_{unl}$ is instantiated using either the \texttt{SimNPO} or \texttt{PROD} loss (Equations~\ref{eq:simnpo-loss},~\ref{eq:prod-loss}).

%% file: src/exp-settings.tex
\subsection{Research Questions}
%%Phan này nếu giới hạn độ dài thì e nghĩ k cần viết cô ạ
\smallskip
\noindent
$\triangleright$ \rqone~%In this RQ, 
We evaluate deprecated API suppression on both the forgetting and a disjoint test set to assess the effectiveness of the unlearning process and the model’s ability to generalize to unseen data.

\smallskip
\noindent
$\triangleright$ \rqtwo~This RQ evaluate whether the model correctly adopts updated APIs to replace deprecated ones in code samples from both forgetting set and unseen test set.

%Beyond reducing deprecated API usage and improving replacement behavior, it is important to verify whether the unlearning process harms the model’s broader programming ability. We therefore use \texttt{HumanEval} as an independent benchmark to assess how well the model preserves its general code generation performance after applying unlearning and contrastive unlearning.

\smallskip
\noindent
$\triangleright$ \rqthree~%In this RQ, 
We evaluate the impact of CURE on the model’s general code generation capability using a widely-used %code generation 
benchmark, i.e., \texttt{HumanEval}~\cite{chen2021evaluating}, examining whether unlearning deprecated APIs degrades overall functional correctness and generation quality.
%Beyond reducing deprecated API usage and improving replacement behavior, it is important to verify whether the unlearning process harms the model’s broader programming ability. We therefore use HumanEval as an independent benchmark to assess how well the model preserves its general code generation performance after applying unlearning and contrastive unlearning.

%, nguyênbảncủaLLM         $\triangleright$ \rqfour

%In our work, we investigate whether the integration of Machine Unlearning and RACG creates a synergistic effect that surpasses the performance of individual methods. Specifically, we examine if weakening a model's internal preference for deprecated APIs through unlearning can resolve the "knowledge conflict" that typically occurs when retrieved documentation contradicts the model's parametric memory. By comparing the combined performance of our unlearning methods (\texttt{SimNPO} and \texttt{PROD}) with RACG against their standalone versions, we assess whether this two-phase approach provides a more robust and precise solution for ensuring the generation of modern, up-to-date code.

\subsection{Dataset Construction}
Following prior work, we adopt the deprecated API evaluation benchmark introduced by Wang et al.~\cite{wang2025llms}, which covers 8 Python libraries with 145 deprecated-to-updated API mappings. The benchmark consists of two datasets: (i) an outdated set containing 9,087 real-world code samples using deprecated APIs, and (ii) an up-to-date set containing 18,340 samples using the corresponding updated APIs. To avoid context leakage, we remove up-to-date samples sharing similar code contexts with those in the outdated set. The remaining 16,423 samples therefore reflect new and unseen contexts in which developers already adopt up-to-date APIs.

%Our data construction is based on the benchmark introduced by Wang et al.~\cite{wang2025llms}. This benchmark provides two groups of code completion samples: $\mathcal{D}_{\text{outdated}}$, which contains real-world usages of deprecated APIs, and $\mathcal{D}_{\text{uptodate}}$, which contains usages of up-to-date or replacement APIs. Based on these two groups, we construct two datasets for each target model: the forget dataset $\mathcal{D}_{\text{forget}}$ and the contrastive dataset $\mathcal{D}_{\text{contrast}}$.

\subsubsection{Forgetting Dataset}
\label{sec:forget-data}
% In traditional machine unlearning methods in NLP, a forget dataset is typically constructed from the training set with the goal of removing undesirable samples. However, our model does not have access to the original training set, so we do not use the original training data. 
% Additionally, if we were to only use the samples generated with deprecated APIs to create \( D_f \), 
% it would only address error signals at a specific point in time, without tackling their root cause
% it would introduce bias, as it would only rectify the model’s error signals at a specific point in time, without addressing the underlying cause of these error signals, thereby lacking generalizability.

% Therefore, we construct \( D_f \) from all the samples in \( D_{\text{outdated}} \) to cover real-world cases of deprecated API usage. In addition, we extract the samples from \( D_{\text{uptodate}} \) where the model still generates deprecated APIs. This behavior indicates that the generation of deprecated APIs is not solely due to outdated contexts but also a result of outdated knowledge encoded within the model's parameters.

Unlike traditional machine unlearning settings, where the forget set is typically sampled from the original training corpus, we do not have access to the pretraining data of Code LLMs. Therefore, we construct the forget set $\mathcal{D}_f$ directly from the outdated set of the benchmark proposed by Wang et al.~\cite{wang2025llms}, which contains real-world deprecated API usages collected from public GitHub repositories. This design enables the model to unlearn deprecated API usage patterns across diverse real-world contexts. 
%In addition, we run the before-unlearning model on contexts from the up-to-date set and identify cases where the model still generates deprecated APIs. These samples are also incorporated into the forget set, as they indicate that deprecated API usage is encoded in the model parameters rather than solely triggered by outdated contexts.
In addition, we randomly sample a limited number of hard negative cases from the up-to-date set where the pre-unlearning model still generates deprecated APIs despite new API contexts. These samples are incorporated into the forget set to expose the model to challenging contexts where deprecated API usage persists despite the absence of outdated contextual cues.

% Hence, for each model, we define the forget dataset as:

% \[
% D_f = D_{\text{outdated}} \cup D_{\text{uptodate} \rightarrow \text{deprecated}},
% \]
% where $\mathcal{D}_{\text{uptodate} \rightarrow \text{deprecated}}$ represents the subset of $\mathcal{D}_{\text{uptodate}}$ for which the model continues to generate deprecated APIs. This subset is constructed separately for each target model, depending on its unique behavior and retained knowledge.

% With this design, we have filtered and generated the sample distribution for each model in the \( D_f \) dataset, as shown in Table \ref{subsec:data}.

\subsubsection{Contrastive Dataset}
\label{sec:contrastive-data}
% The contrastive dataset \( \mathcal{D}_{\text{contrast}} \) is constructed to support the replacement-oriented objective introduced in Section~\ref{sec:contrastive}. We construct \( \mathcal{D}_{\text{contrast}} \) from samples in \( \mathcal{D}_{\text{outdated}} \) and \( \mathcal{D}_{\text{uptodate}} \) where deprecated-to-replacement API pairs can be identified. For samples from \( \mathcal{D}_{\text{uptodate}} \), the available up-to-date API usage is used to form the positive completion. For samples from \( \mathcal{D}_{\text{outdated}} \) that do not contain a ready-to-use replacement completion, we use \texttt{Gemini-2.5-Pro} to generate the positive completion based on the known replacement API while preserving the original code intent.
% This construction of \( \mathcal{D}_{\text{contrast}} \) allows the model not only to suppress deprecated APIs but also to prefer modern, valid alternatives. By incorporating both the negative samples (deprecated APIs) and positive samples (replacement APIs), the dataset guides the model toward more reliable adaptation to the evolving software ecosystem.
To promote correct API replacement, we construct a contrastive dataset $\mathcal{D}_{\text{ctr}}$ consisting of negative samples that contain deprecated API usages generated by the model. 
For outdated-context samples, the corresponding positive samples are generated using \texttt{Gemini-2.5-Pro}, rewriting the code with the correct replacement API while preserving the original intent and context.
In addition, for the selected up-to-date contexts described earlier where the model still generates deprecated APIs, we directly use the corresponding ground-truth up-to-date code %samples 
as positive examples. 
%This construction enables the model not only to suppress deprecated APIs but also to prefer correct modern replacements.

% With this design, we report the number of contrastive samples for each model in Table \ref{subsec:data}.

\subsubsection{Testing Dataset}
\label{sec:test-data}
% We evaluate each model under two test settings to measure both direct forgetting and generalization. The first evaluation is conducted on \( \mathcal{D}_{f} \), which is used to assess how effectively the model suppresses the deprecated API behaviors targeted during unlearning. This setting allows us to examine whether the updated model not only reduces deprecated API generation, but also redirects its outputs toward the corresponding replacement APIs.
% The second evaluation is conducted on \( \mathcal{D}_{\text{uptodate}} \), which serves as an unseen test set for assessing generalization. Unlike \( \mathcal{D}_{f} \), this dataset is not used as the direct unlearning target. It therefore allows us to evaluate whether the model can transfer the learned replacement-oriented behavior to new code contexts and generate up-to-date API usages beyond the samples involved in the unlearning process.
The testing set consists of unseen code contexts used to evaluate the generalization capability of the model after unlearning. Specifically, we construct this set from the up-to-date dataset of the benchmark proposed by Wang et al.~\cite{wang2025llms} after removing all samples involved in ($\mathcal{D}_{f}$). This setup enables us to assess whether the model can generalize replacement-oriented behavior to unseen contexts and correctly generate up-to-date API usages beyond the samples used during unlearning.

\begin{table}
\centering
\caption{Dataset statistics of  $\mathcal{D}_{f}$, $\mathcal{D}_{\text{ctr}}$ and $\mathcal{D}_{\text{test}}$ across 4 Code LLMs.}
\begin{tabular}{l|c|c|c}
\hline
\textbf{Model}  & $\mathcal{D}_{f}$ & $\mathcal{D}_{\text{ctr}}$ & $\mathcal{D}_{\text{test}}$ \\
\hline
\texttt{DeepSeek-Coder-1.3B}   & 9,436 & 3,440 & 16,376 \\
%\hline
\texttt{StarCoder2-3B}         & 11,050 & 6,919 & 14,865  \\
%\hline
\texttt{CodeLlama-7B}         & 10,396 & 7,377 & 15,374 \\
%\hline
\texttt{CodeGen-2B}           & 10,801 & 7,662 & 14,832 \\
\hline
\end{tabular}
\label{tab:data}
\end{table}

In this work, we evaluate four Code LLMs: \DeepSeek, \StarCoder, \CodeLlama and \CodeGenBase. For each model, we construct the corresponding datasets  $\mathcal{D}_f$, $\mathcal{D}_{\text{ctr}}$ and $\mathcal{D}_{\text{test}}$.
Table~\ref{tab:data} presents the statistics of the constructed datasets in terms of the number of samples.

\subsection{Evaluation Metrics}
\label{sec:metrics}
We evaluate unlearning effectiveness using three metrics: Deprecated API Usage Rate (DUR), Replacement API Usage Rate (RUR), and Mismatch API Usage Rate (MUR)~\cite{Wang2024}, which respectively measure the proportion of outputs containing the target deprecated API, the corresponding replacement API, or neither of them. 
In particular, MUR captures cases where the model successfully avoids the deprecated API but fails to generate the intended valid replacement, instead producing a different API that does not correspond to the intended valid replacement. 
API usage is identified through alias-based regular-expression matching with word-boundary constraints to reduce false positives~\cite{wang2025llms}.

%These metrics indicate whether the model generates the deprecated API, produces the corresponding replacement API, or outputs neither of them.

% API usage is identified using the alias dictionary associated with each sample. For each target API, we match its corresponding aliases against the generated code using regular expressions with word-boundary constraints, rather than simple substring matching. This reduces false positives caused by partial overlaps with variable names, strings, or other API names. The metrics are defined as follows:
% \begin{itemize}
% \item \textbf{DUR}: proportion of outputs containing the target deprecated API.
% \item \textbf{RUR}: proportion of outputs containing the corresponding replacement API.
% \item \textbf{MUR}: proportion of outputs containing neither the deprecated nor replacement API.
% \end{itemize}
For RQ$_3$, we evaluate the model's general %code 
generation capability using \texttt{Pass@k} on the HumanEval benchmark~\cite{chen2021evaluating}. \texttt{Pass@k} measures the probability that at least one out of $k$ generated %code 
samples correctly solves a given %programming 
task. Following prior work~\cite{du2024evaluating,yang2024evaluation,jiang2026survey}, we report results with $k = 1,3,5$.

\subsection{Implementation}
\label{sec:implementation}
All experiments were conducted on a single NVIDIA A100 GPU with 80GB VRAM. Input prompts were truncated to a maximum context length of 2,048 tokens, while generation outputs were limited to 128 tokens, which is sufficient for API transformation tasks without producing unnecessarily verbose code. We set $\beta = 0.1$ for the \texttt{SimNPO} loss and $\lambda = 1$ to control the contribution of the contrastive objective. %During inference, we used a temperature of $0$ and a fixed random seed of $42$ to ensure deterministic and reproducible results. 

% across all experiments.
%All experiments, including machine unlearning, retrieval-augmented inference, and metric evaluation, were conducted on a single NVIDIA A100 GPU with 80GB of VRAM. 
%Our framework is built upon \DeepSeek \StarCoder \CodeLlama \CodeGenBase as the pretrained backbone. 
%To balance computational efficiency and numerical stability, all model parameters are loaded and evaluated in \texttt{bfloat16} precision, which reduces memory usage while maintaining stable performance. 

%Input prompts are constrained to a maximum context length of 2{,}048 tokens to align with the model’s architectural limits. 
%Generated outputs are capped at 128 new tokens, which is sufficient for completing API-level code transformations while avoiding unnecessarily long or verbose generations. 

%During inference, we adopt a deterministic decoding strategy to ensure consistent and comparable behavior across all models. Specifically, the temperature is set to 0, forcing the model to select the most probable token at each step. This configuration eliminates stochastic variability and allows for a fair comparison of model behaviors under different settings. In addition, we fix the random seed to 42 to ensure full reproducibility across all runs.

%% file: src/results.tex
\subsection{\rqone}

\begin{figure}
    \centering
    \includegraphics[width=0.96\linewidth]{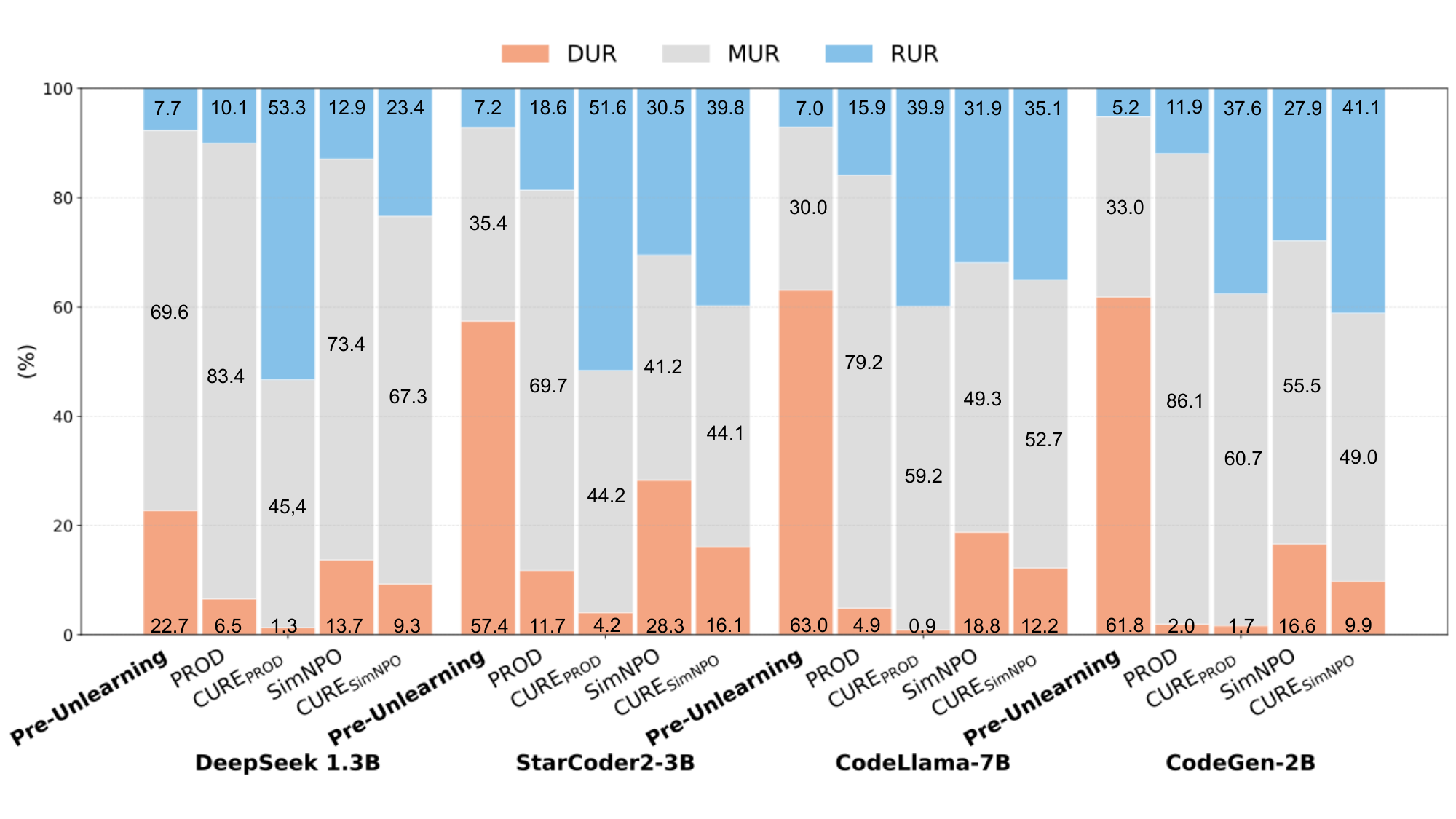}
    \caption{Generation Behavior Distribution on $\mathcal{D}_f$ across Unlearning Methods.}
    %Relative DUR Reduction (\%) Compared to Before-Unlearning Baselines
    \label{fig:df-stackbar}
\end{figure}

\begin{figure}
    \centering
    \includegraphics[width=0.96\linewidth]{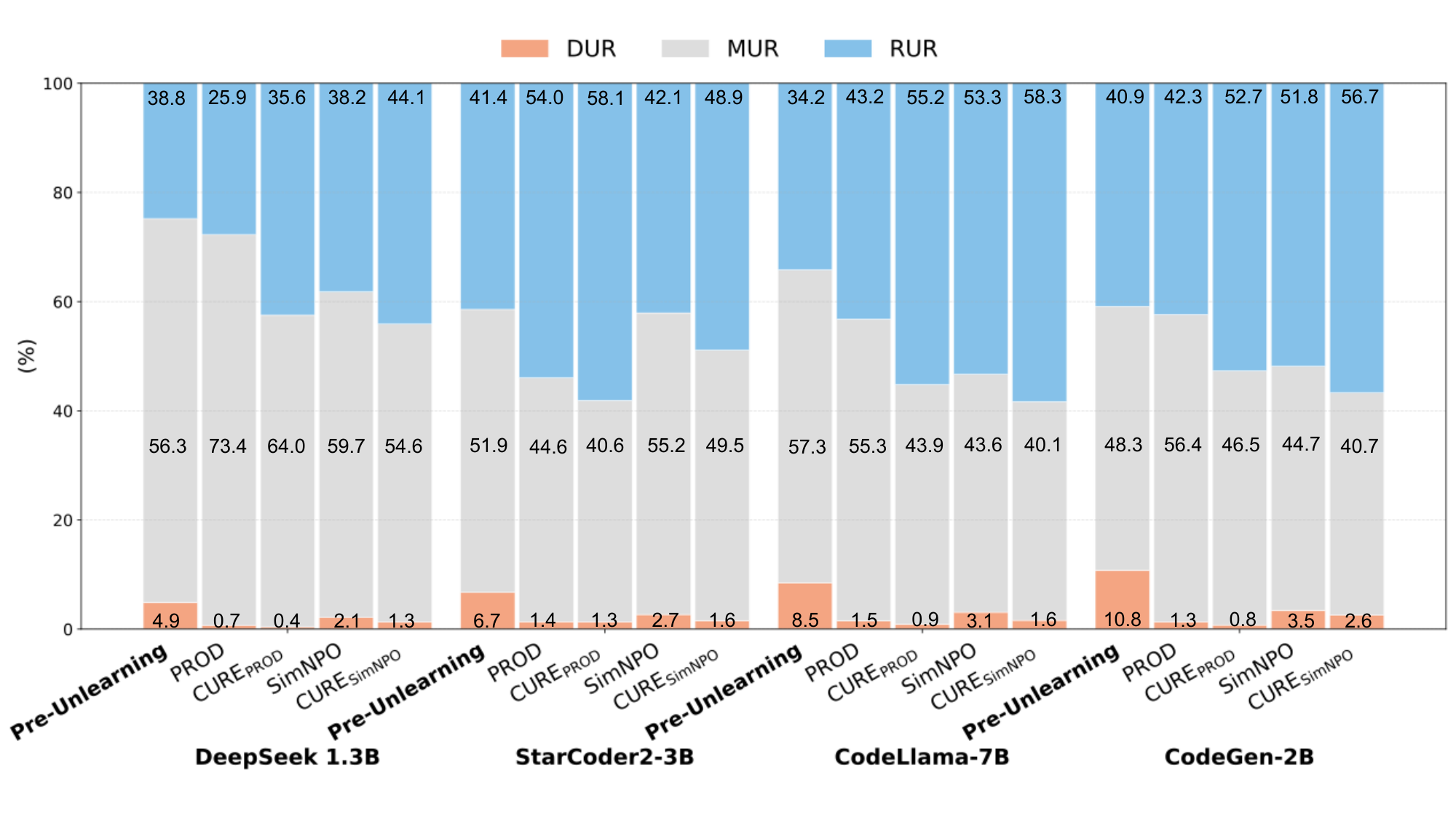}
    \caption{Generation Behavior Distribution on Up-to-date Contexts $\mathcal{D}_{\text{test}}$.}
    %Relative DUR Reduction (\%) Compared to Before-Unlearning Baselines
    \label{fig:dtest-stackbar}
\end{figure}

\begin{table*}[t]
\centering
\caption{Comparison of Model Utility on \texttt{HumanEval} Across Unlearning Methods.}
\label{tab:base-model-comparison}
\resizebox{\textwidth}{!}{
\begin{tabular}{lcccccccccccc}
\toprule
\multirow{2}{*}{\textbf{Method}} 
& \multicolumn{3}{c}{\textbf{\DeepSeek}} 
& \multicolumn{3}{c}{\textbf{\StarCoder}} 
& \multicolumn{3}{c}{\textbf{\CodeLlama}} 
& \multicolumn{3}{c}{\textbf{\CodeGenBase}} \\
\cmidrule(lr){2-4} \cmidrule(lr){5-7} \cmidrule(lr){8-10} \cmidrule(lr){11-13}
& \textbf{Pass@1} & \textbf{Pass@3} & \textbf{Pass@5}
& \textbf{Pass@1} & \textbf{Pass@3} & \textbf{Pass@5}
& \textbf{Pass@1} & \textbf{Pass@3} & \textbf{Pass@5}
& \textbf{Pass@1} & \textbf{Pass@3} & \textbf{Pass@5} \\
\midrule
\textbf{\texttt{Pre-Unlearning}}
& {41.76}        & {49.52}     & {51.78}
& {18.03}             & {31.43}     & {36.82}
& {29.88}             & {38.72}     & {43.90}
& {24.88}             & {30.00}     & 31.71 \\

\textbf{\texttt{PROD}~\cite{jiang2026large}} 
& 38.66 & 47.56 & 50.17
& 14.51 & 22.83 & 28.34
& 25.38 & 35.18 & 39.02
& 19.99 & 25.21 & 30.32 \\

\textbf{$\texttt{CURE}_{\texttt{PROD}}$} 
& {\bf 40.37} & {{\bf50.43}} & {{\bf54.27}}
& {\bf15.23} & \underline{25.54} & \underline{31.43}
& {\bf27.32} & {\bf38.11} & {\bf43.86}
& \underline{22.39} & \underline{29.01} & \underline{32.34} \\

\textbf{\texttt{SimNPO}~\cite{fan2024simplicity}} 
& 39.65 & 46.71 & 48.78
& 13.37 & 23.77 & 28.90
& 25.01 & 35.92 & 39.56
& 20.73 & {28.48} & {30.32} \\

\textbf{$\texttt{CURE}_{\texttt{SimNPO}}$} 
& \underline{39.78} & \underline{48.49} & \underline{50.83}
& \underline{14.94} & \underline{\bf26.88} & \underline{\bf32.05}
& \underline{26.10} & \underline{36.44} & \underline{40.41}
& {\bf23.08} & {\bf30.18} & {\bf32.96} \\
\bottomrule
\end{tabular}
}
\vspace{4pt}

\parbox{\linewidth}{\scriptsize\itshape
Note: The best-performing unlearning approach in each column is shown in {\bf bold}, while the second-best result is \underline{underlined}.
}

\end{table*}

We measure three metrics DUR, RUR and MUR of all studied models on both $\mathcal{D}_f$ and $\mathcal{D}_{\text{test}}$ under three settings: (i) pre-unlearning, (ii) after applying traditional unlearning (i.e., \texttt{PROD} and \texttt{SimNPO}), and (iii) after applying the proposed contrastive unlearning (i.e., $\texttt{CURE}_{\texttt{PROD}}$ and $\texttt{CURE}_{\texttt{SimNPO}}$). 
Evaluation on $\mathcal{D}_f$ allows us to directly assess whether the unlearning process successfully suppresses deprecated API usage on the targeted forgetting samples, while evaluation on $\mathcal{D}_{\text{test}}$ examines whether the learned behavior generalizes to up-to-date code contexts.
As shown in Figs.~\ref{fig:df-stackbar} and~\ref{fig:dtest-stackbar}, the proportion of deprecated API usages (orange bars) consistently decreases across all unlearning methods on both $\mathcal{D}_f$ and $\mathcal{D}_{\text{test}}$ compared to the original pre-unlearning models, confirming the effectiveness of parametric unlearning in mitigating outdated API behaviors.

On $\mathcal{D}_f$, the %original 
models exhibit high DUR, ranging from 22.7\% to 63.0\%. Both \texttt{PROD} and \texttt{SimNPO} substantially suppress deprecated API usage, with \texttt{PROD} consistently outperforming \texttt{SimNPO}, e.g., %For instance, 
\texttt{PROD} achieves up to a 96.8\% relative DUR reduction, whereas \texttt{SimNPO} gets at most 73.1\% on \CodeGenBase, suggesting that token-level probability redistribution is more effective than sequence-level suppression. 
Furthermore, incorporating the proposed contrastive objective further improves deprecated API suppression across most settings. For example, compared to \texttt{PROD}, $\texttt{CURE}_{\texttt{PROD}}$ %further 
reduces DUR 
%from 6.5\% to 1.3\% on \DeepSeek and 
from 11.7\% to 4.2\% on \StarCoder. 
Similar improvements are %also 
observed for $\texttt{CURE}_{\texttt{SimNPO}}$, indicating that explicitly contrasting deprecated APIs with valid alternatives enhances forgetting effectiveness.
On %the test set 
$\mathcal{D}_{\text{test}}$, the original models already show relatively low DUR prior to unlearning, ranging from 4.9\% to 10.8\%, indicating that deprecated APIs are less frequently activated in up-to-date  contexts. 
Nevertheless, all unlearning methods consistently maintain lower DUR than the original models, demonstrating that the suppression effect generalizes beyond the forgetting samples.

%\vspace{.1cm}
\begin{shadedbox}
	\textbf{Answer to RQ$_1$.} All unlearning methods reduce deprecated API usage, while the proposed \CURE variants consistently achieve stronger and more generalizable mitigations.

\end{shadedbox}

\subsection{\rqtwo}

% \begin{table}[t]
% \centering
% \setlength{\tabcolsep}{2.5pt}
% \caption{}
% \label{tab:api-generation-results-scaled}
% \begin{tabular}{lcccccccccccc}
% \toprule
% \multirow{2}{*}{\textbf{Method}} 
% & \multicolumn{3}{c}{\textbf{\texttt{DS}}} 
% & \multicolumn{3}{c}{\textbf{\texttt{SC}}} 
% & \multicolumn{3}{c}{\textbf{\texttt{CL}}} 
% & \multicolumn{3}{c}{\textbf{\texttt{CG}}} \\
% \cmidrule(lr){2-4} \cmidrule(lr){5-7} \cmidrule(lr){8-10} \cmidrule(lr){11-13}
% & \textbf{R}$\uparrow$ & \textbf{D}$\downarrow$ & \textbf{M}$\downarrow$
% & \textbf{R}$\uparrow$ & \textbf{D}$\downarrow$ & \textbf{M}$\downarrow$
% & \textbf{R}$\uparrow$ & \textbf{D}$\downarrow$ & \textbf{M}$\downarrow$
% & \textbf{R}$\uparrow$ & \textbf{D}$\downarrow$ & \textbf{M}$\downarrow$ \\
% \midrule

% \textbf{\texttt{PROD}} 
% & 11.1 & 8.9 & 80.0
% & 10.7 & 10.3 & 79.0
% & 11.1 & 16.3 & 72.6
% & 13.4 & 12.1 & 74.5 \\

% \textbf{$\texttt{CURE}_{\texttt{P}}$}
% & 56.6 & 6.3 & 37.1
% & 42.3 & 7.7 & 50.0
% & 42.9 & 6.3 & 50.8
% & 34.6 & 5.4 & 60.0 \\

% \midrule

% \textbf{\texttt{SimNPO}}
% & 24.8 & 21.4 & 53.8
% & 20.5 & 20.1 & 59.4
% & 23.2 & 29.9 & 46.9
% & 18.8 & 28.4 & 52.8 \\

% \textbf{$\texttt{CURE}_{\texttt{S}}$}
% & 36.2 & 15.4 & 48.4
% & 30.2 & 15.3 & 54.5
% & 34.7 & 20.8 & 44.5
% & 32.6 & 19.5 & 47.9 \\

% \bottomrule
% \end{tabular}

% \vspace{4pt}

% \parbox{\linewidth}{\scriptsize\itshape
% DS: DeepSeek-1.3B;
% SC: StarCoder2-3B;
% CL: CodeLlama-7B;
% CG: CodeGen-2B.
% }
% \end{table}

On $\mathcal{D}_f$, suppression-oriented unlearning methods, including \texttt{PROD} and \texttt{SimNPO}, consistently reduce deprecated API usage (DUR). However, though these methods partially improve replacement API usage (RUR), they also increase mismatch generations (MUR). For example, on \StarCoder, \texttt{PROD} reduces DUR from 57.4\% to 11.7\% and increases RUR from 7.2\% to 18.6\%, but simultaneously raises MUR from 35.4\% to 69.7\%. Similar trends are observed across other models and unlearning methods, suggesting that suppression-oriented unlearning mainly teaches the model to avoid deprecated APIs, but remains less effective at guiding the model toward semantically correct replacement APIs. %under the same context.
In contrast, incorporating the contrastive unlearning objective %consistently 
improves RUR while reducing MUR across all evaluated models. For instance, compared to \texttt{PROD}, $\texttt{CURE}_{\texttt{PROD}}$ improves RUR from 
%7.2\% to 53.3\% on \DeepSeek, 
18.6\% to 51.6\% on \StarCoder, and 11.9\% to 37.6\% on \CodeGenBase. 
Similar improvements are also observed when comparing \texttt{SimNPO} with $\texttt{CURE}_{\texttt{SimNPO}}$, indicating that contrastive unlearning not only reduces deprecated APIs, but also better guides the model toward generating correct replacement APIs under the same context.

On $\mathcal{D}_{\text{test}}$, the original models already exhibit %relatively 
low deprecated API usage, suggesting that Code LLMs can partially adapt to new contexts and naturally prefer more up-to-date APIs. 
Nevertheless, all unlearning methods %further 
reduce deprecated API generation while improving correct replacement API usage. 
%More importantly, 
Incorporating the proposed contrastive objective consistently yields larger RUR and lower mismatch generations compared to suppression-based methods. For example, on \StarCoder, \texttt{PROD} achieves a relative RUR improvement of 30.4\% over the original model, while $\texttt{CURE}_{\texttt{PROD}}$ further increases it to 40.3\%. 
Similarly, on \CodeLlama, \texttt{SimNPO} achieves a 55.8\% relative RUR improvement, whereas $\texttt{CURE}_{\texttt{SimNPO}}$ reaches 70.5\%. 
% while also reducing MUR from 43.6% to 40.1%. 
%These results suggest that contrastive unlearning not only suppresses deprecated APIs in memorized contexts, but also better generalizes to unseen contexts by steering the model toward semantically correct and up-to-date API usage.
%\vspace{.1cm}
\begin{shadedbox}
	\textbf{Answer to RQ$_2$.} The proposed contrastive unlearning not only reduces deprecated APIs, but also improves the model’s tendency to generate updated APIs while reducing mismatched API usages on up-to-date contexts.
\end{shadedbox}

\subsection{\rqthree}
We evaluate the impact of unlearning on \texttt{HumanEval}~\cite{chen2021evaluating}. %, a widely-used benchmark for assessing the functional code generation capability of Code LLMs. 
Table~\ref{tab:base-model-comparison} reports the $Pass@1$, $Pass@3$, and $Pass@5$ results across four studied models under pre- and post-unlearning settings. Overall, all unlearning approaches introduce only minor degradation on \texttt{HumanEval}.
For example, on \DeepSeek, \texttt{PROD} and \texttt{SimNPO} retain 95.2\% and 94.5\% of the original performance on average, while $\texttt{CURE}_{\texttt{PROD}}$ and $\texttt{CURE}_{\texttt{SimNPO}}$ achieve even stronger preservation rates of 101.1\% and 97.1\%, respectively.
This trend remains consistent across the other models, where the best and second-best results are predominantly achieved by the \texttt{CURE} variants.

\begin{shadedbox}
	\textbf{Answer to RQ$_3$.} %These findings suggest that 
    Explicitly promoting valid replacement APIs preserves general model utility more effectively than suppression-based unlearning approaches.
\end{shadedbox}

%% file: src/threats.tex
{\bf Construct Validity.} Our evaluation relies on alias-aware regular-expression matching to identify deprecated and replacement API usages. Though complex generation patterns or implicit API wrappers may introduce minor matching inaccuracies, we mitigate this threat by leveraging benchmark-provided API alias mappings together with word-boundary constraints. % to reduce false positives. 
Moreover, positive samples in the contrastive dataset are partially generated using \texttt{Gemini-2.5-Pro}, which may introduce generation bias or imperfect transformations. To mitigate this threat, we constrain the generation process to preserve the original code intent and only replace the target APIs while keeping surrounding contexts unchanged.

{\bf Internal Validity.} A potential threat arises from augmenting the forget set with hard negative cases sampled from up-to-date contexts where the baseline model still generates deprecated APIs. It is possible to argue that such samples implicitly expose the model to modern API usages rather than purely removing outdated knowledge. To reduce this threat, we only incorporate a limited number of randomly sampled hard negatives and use them as robustness-oriented failure cases instead of direct replacement supervision.